\documentclass[sigconf,authorversion]{acmart}

\usepackage{booktabs} 

\setcopyright{rightsretained}

\setcitestyle{square}

\acmDOI{00.0005/123_4}

\acmISBN{123-4567-00-00/00/00}

\acmConference[SIGGRAPH Asia 2017 Posters]{SIGGRAPH Asia 2017 Posters}{November 2017}{Bangkok, Thailand}
\acmYear{2017}
\copyrightyear{2017}
\acmPrice{00.00}

\begin{document}
\title{Visualization and Labeling of Point Clouds in Virtual Reality}

\author{Jonathan Dyssel Stets}
\affiliation{%
  \institution{Technical University of Denmark}
}

\author{Yongbin Sun}
\affiliation{%
  \institution{Massachusetts Institute of Technology}
}

\author{Wiley Corning}
\affiliation{%
  \institution{MIT Media Lab}
}

\author{Scott Greenwald}
\affiliation{%
  \institution{MIT Media Lab}
}

\renewcommand{\shortauthors}{Stets et. al.}

%
%




\begin{teaserfigure}
\begin{tabular}{@{}c@{}c@{}}
	 \includegraphics[width = 0.5\textwidth]{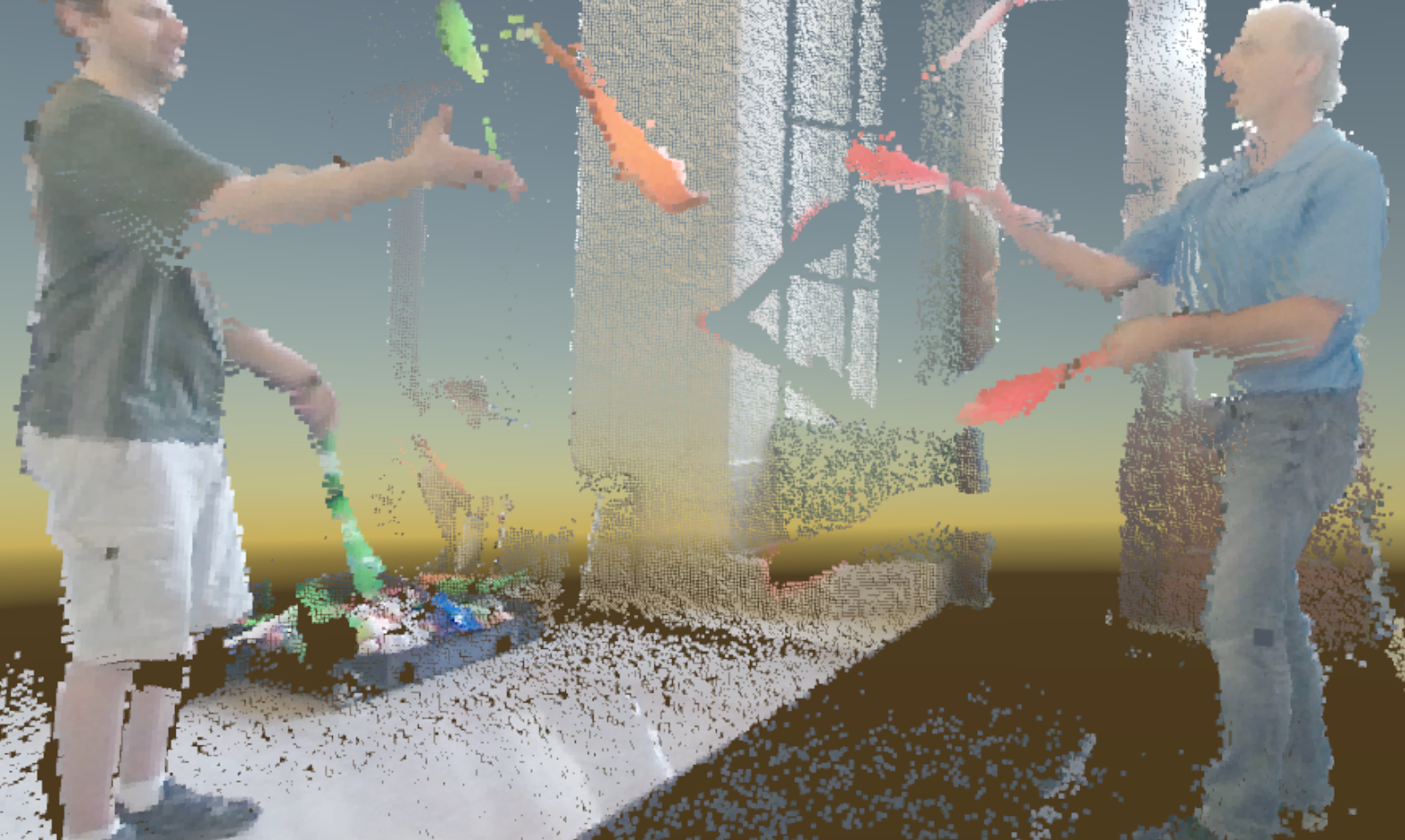} &
\hspace{2pt}
		 \includegraphics[width = 0.5\textwidth]{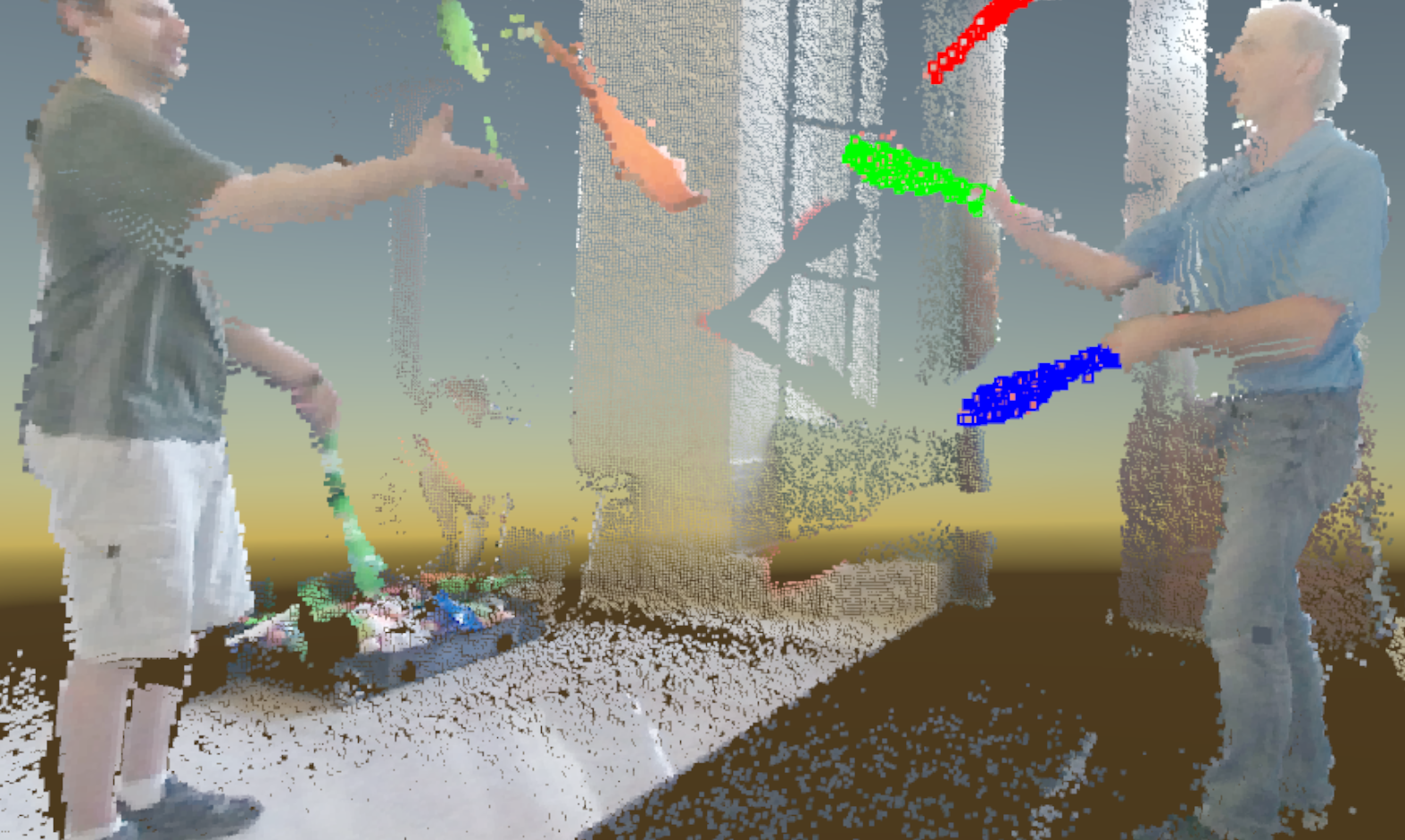}  \\[-2ex]
\end{tabular}
  \caption{Screen capture from the Virtual Reality application. A point cloud of jugglers is displayed, the left and right images shows the point cloud before and after the juggling pins has been annotated by the user.}
  \label{fig:teaserimg}
\end{teaserfigure}

\maketitle
\section{Introduction}
We present a Virtual Reality (VR) application for labeling and handling point cloud data sets. 
A series of room-scale point clouds are recorded as a video sequence using a Microsoft Kinect. 
The data can be played and paused, and frames can be skipped just like in a video player.
The user can walk around and inspect the data while it is playing or paused.
Using the tracked hand-held controller, the user can select and label individual parts of the point cloud. 
The points are highlighted with a color when they are labeled. 
With a tracking algorithm, the labeled points can be tracked from frame to frame to ease the labeling process.
Our sample data is an RGB point cloud recording of two people juggling with pins.
Here, the user can select and label, for example, the juggler pins as shown in Figure \ref{fig:teaserimg}. Each juggler pin is labeled with various colors to indicate different labels.

\section{Motivation}
We consider the use case of viewing animated point cloud data in VR. Although individual frames appear grainy and low-resolution, when they are animated, they produce a lifelike effect that is suitable for representing dynamic real-world scenes. However, without any labeling or metadata, it is diffcult to edit or modify these scenes. Labeling a person with a name, or changing the color of an object would be examples of basic editing tasks. Segmenting the point clouds in each frame, and establishing how the segments in each frame correspond with those in surrounding frames, makes it possible to perform such tasks.

Furthermore, the same set of user interface affordances can be used for other use cases as well. The demand for large labeled point cloud data sets is increasing as the need for classification algorithms for such data is gaining popularity, especially for data-driven approaches like deep neural networks. Navigation of self driving cars, social robots and robot interaction in general often rely on 3D data that needs to be classfied or segmented. Additionally, the use of 3D sensors is increasing, while these sensors are getting more mobile, better and more available. This enables the potential for more and better 3D point cloud data. The website Semantic3D \cite{semantic3d} provides examples of large labeled 3D point cloud data sets. Oftentimes, such 3D data sets are labeled manually using a 2D computer monitor and a computer mouse or similar input device. We argue that a two-dimensional human-computer interaction with three-dimensional data can be inefficient in many cases. For example, the labeling problem is more difficult with a 2D interface when the orthogonal $xy$, $xz$, or $yz$ planes do not separate the points to be labeled from the others surrounding them.

We present an application that demonstrates an intuitive way to inspect and label 3D point cloud data. The user simply inspects the data by moving around in the 3D point cloud, as one would do in the real world. Labeling is done by pointing and drawing in the 3D space with a hand-held controller to mark relevant parts of the data.

\section{Details}
\begin{figure}
	 \includegraphics[width = 0.8\columnwidth]{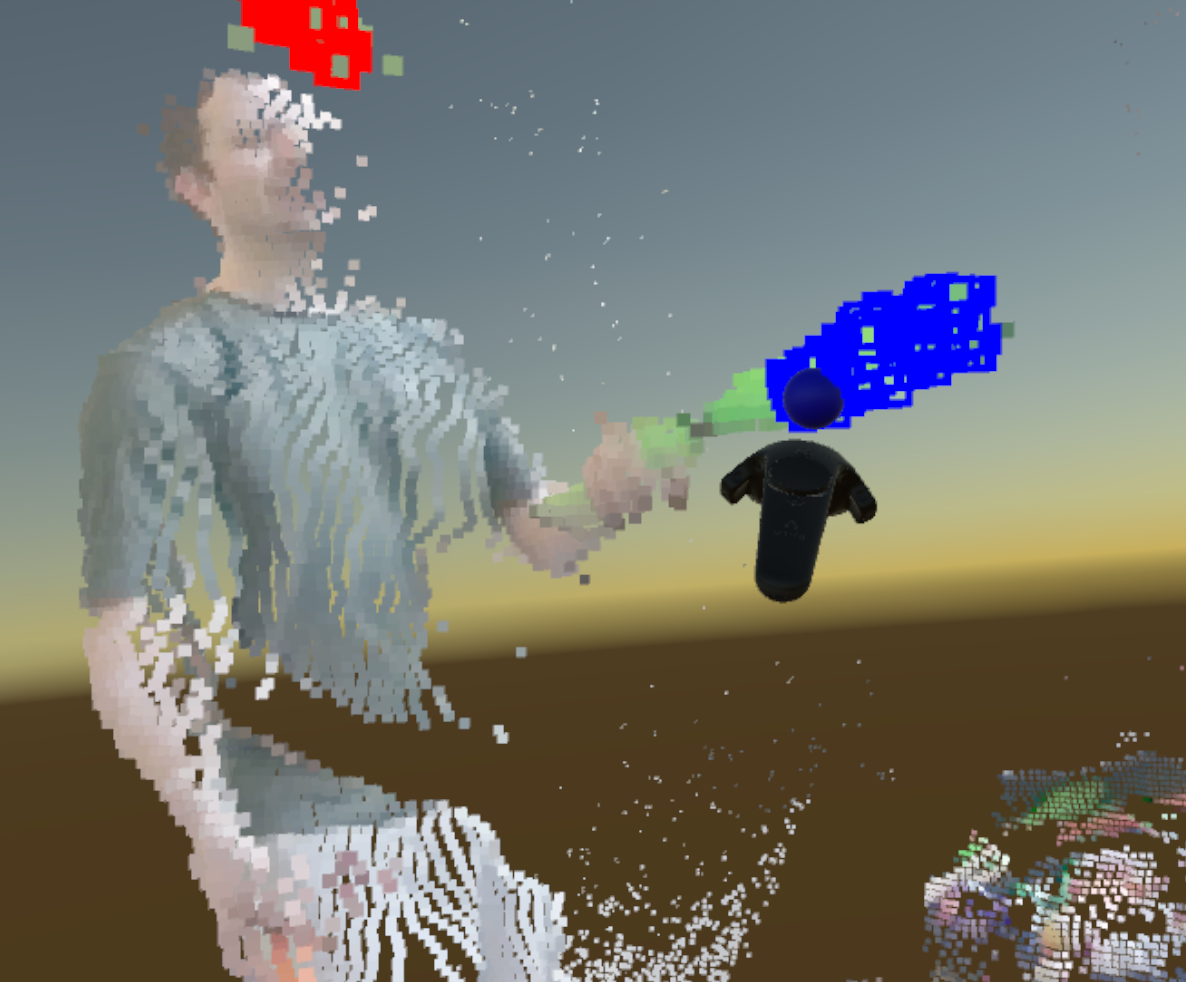}
  \caption{The juggling pin is being labeled with a blue label using the hand-held controller. Haptic feedback in the con- troller is provided when points are selected.}
  \label{fig:controller}
\end{figure}

We have recorded a data set using a Microsoft Kinect 2 and a custom- built software tool. The software records and converts the depth map and rgb video stream from the Kinect to a point cloud and stores the $(X,Y,Z,R,G,B)$  data in a binary format. The point clouds are recorded and stored approximately 30 times pr. seconds. The set of point clouds is loaded into the VR application and visualized at room scale. We use the HTC Vive setup for VR, which enables tracking of both headset and the hand-held controllers. Using the touchpad on the hand-held controller, the user can navigate back and forward in the point cloud frames. A marker-tool is attached to the controller and this can be used like a paint brush to label points in the point cloud (see Figure \ref{fig:controller}). Haptic feedback is provided when selecting the points to add the feeling of physically interacting with the data. Individual parts of the point cloud can be labeled with different labels, each represented by varying colors. The user can switch between labels on the controller touchpad, as well as choosing an eraser-tool to unlabel points in the point cloud. We argue that interacting with 3-dimensional data in this way is an easy and fast alternative to a traditional 2-dimensional interface such as a mouse or a touch screen.

\begin{figure}
\begin{tabular}{@{}c@{}c@{}c@{}}
	 \includegraphics[width = 0.33\linewidth]{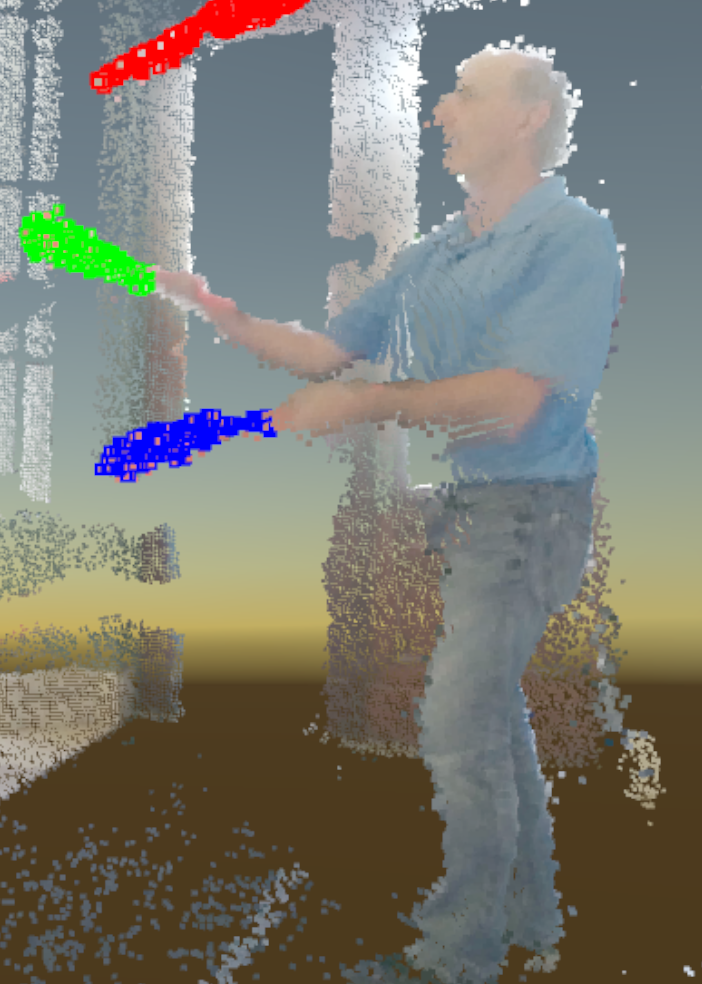} &
\hspace{2pt}
\includegraphics[width = 0.33\linewidth]{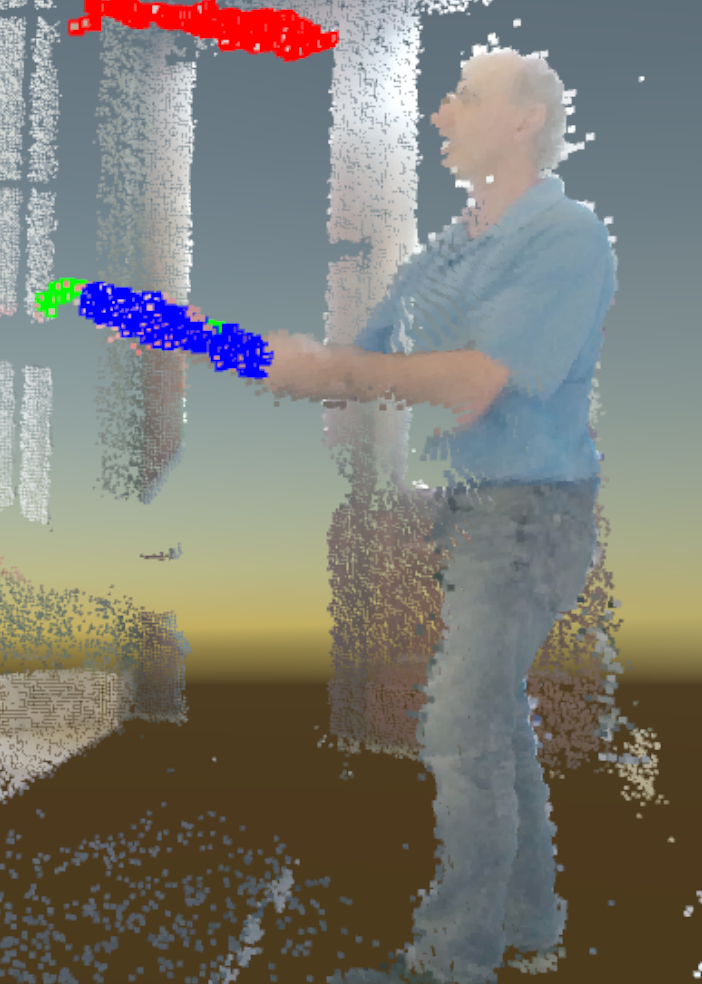} &
\hspace{2pt}
		 \includegraphics[width = 0.33\linewidth]{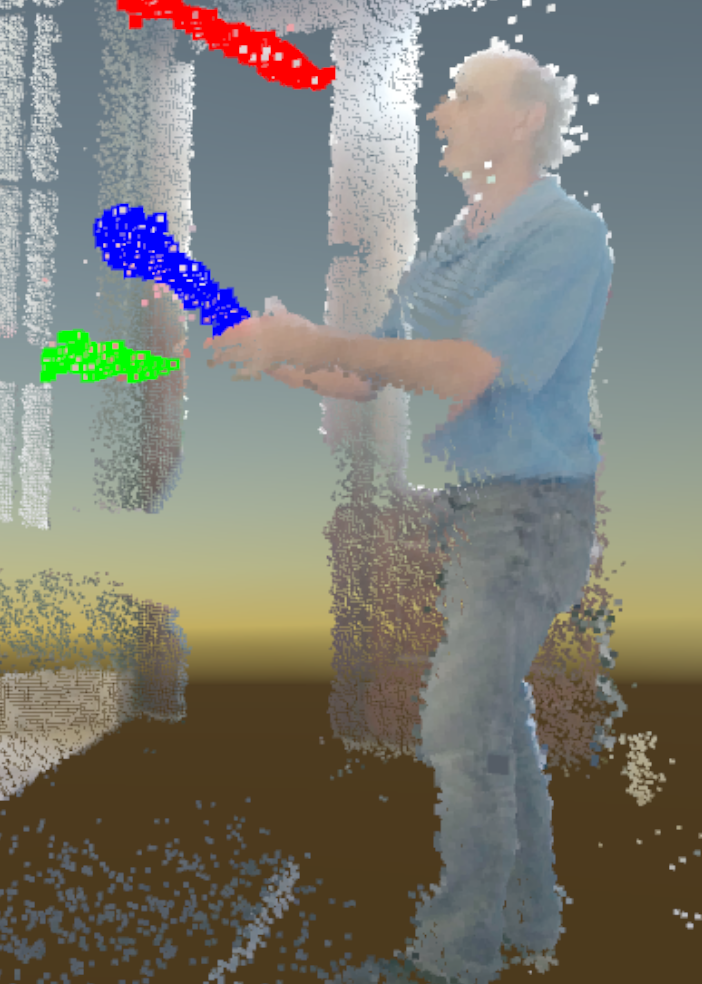}  \\[-2ex]
\end{tabular}
  \caption{Three frames from the point cloud data. The juggling pins are labeled individually (red, green and blue) by the user in the  first frame. The tracking algorithm is used to track points from one frame and automatically label them in the next.}
  \label{fig:icp}
\end{figure}

To ease the labeling across frames, we have added a tracking feature to automatically label subsequent frames in the point cloud set. The tracking algorithm fits labeled points in $frame_i$ to a portion of points in $frame_{i+1}$, so that we can predict labels from one frame to the next. Figure \ref{fig:icp} shows how three individual parts of the point cloud has been labeled, and then tracked from one frame to the next. We implemented two algorithms to fit labeled points. The first one is an Iterative Closest Point (ICP) algorithm \cite{besl1992method}, which iteratively detects the closest point in $frame_{i+1}$
labeled point in $frame_i$ and estimates the current best rotation and translation to update the positions of the labeled points. This iterative process terminates until a predetermined stop criteria satisfies, for example, the spatial difference between labeled points and matched points converges. The second algorithm follows the same pipeline as the first one, but is augmented with color support to improve tracking accuracy. Instead of detecting the spatially closest point, the second algorithm selects, for each query point, the point with the most similar color from k-nearest neighboring points as the match. We build a k-d tree \cite{brown2014building}, which can be pre-calculated for each point cloud, to expedite the nearest neighbor search.

Our method is not limited to Kinect data, thus data of varying scale and resolution can be visualized and labeled. We also emphasize the potential of this application for modifying point clouds, such as deleting noisy points, or cutting out relevant parts of the data using the same type of labeling and selection method.

\section{Conclusion}
We have realized an application for visualizing and labeling point clouds in Virtual Reality. The application demonstrates that it is possible to label and inspect large sets of room-scale point clouds in a fast and intuitive way. With the application, it is possible to select and track parts of the data to speed up labeling in sequences of point clouds. We believe that is a step in the direction towards incorporating point clouds into an e cient production pipeline for realistic animated virtual reality content, and more high precision labeled point cloud data sets for a variety of other applications as well.

\section{Acknowledgements}
Thanks to the MIT Juggling Club for allowing us to record the data.

\bibliographystyle{ACM-Reference-Format}
\bibliography{references}{}


\begin{thebibliography}{00}


\ifx \showCODEN    \undefined \def \showCODEN     #1{\unskip}     \fi
\ifx \showDOI      \undefined \def \showDOI       #1{#1}\fi
\ifx \showISBNx    \undefined \def \showISBNx     #1{\unskip}     \fi
\ifx \showISBNxiii \undefined \def \showISBNxiii  #1{\unskip}     \fi
\ifx \showISSN     \undefined \def \showISSN      #1{\unskip}     \fi
\ifx \showLCCN     \undefined \def \showLCCN      #1{\unskip}     \fi
\ifx \shownote     \undefined \def \shownote      #1{#1}          \fi
\ifx \showarticletitle \undefined \def \showarticletitle #1{#1}   \fi
\ifx \showURL      \undefined \def \showURL       {\relax}        \fi
\providecommand\bibfield[2]{#2}
\providecommand\bibinfo[2]{#2}
\providecommand\natexlab[1]{#1}
\providecommand\showeprint[2][]{arXiv:#2}

\bibitem[\protect\citeauthoryear{Besl, McKay, et~al\mbox{.}}{Besl
  et~al\mbox{.}}{1992}]%
        {besl1992method}
\bibfield{author}{\bibinfo{person}{Paul~J Besl}, \bibinfo{person}{Neil~D
  McKay}, {et~al\mbox{.}}} \bibinfo{year}{1992}\natexlab{}.
\newblock \showarticletitle{A method for registration of 3-D shapes}.
\newblock \bibinfo{journal}{{\em IEEE Transactions on pattern analysis and
  machine intelligence\/}} \bibinfo{volume}{14}, \bibinfo{number}{2}
  (\bibinfo{year}{1992}), \bibinfo{pages}{239--256}.
\newblock


\bibitem[\protect\citeauthoryear{Brown}{Brown}{2014}]%
        {brown2014building}
\bibfield{author}{\bibinfo{person}{Russell~A Brown}.}
  \bibinfo{year}{2014}\natexlab{}.
\newblock \showarticletitle{Building a balanced kd tree in o (kn log n) time}.
\newblock \bibinfo{journal}{{\em arXiv preprint arXiv:1410.5420\/}}
  (\bibinfo{year}{2014}).
\newblock


\bibitem[\protect\citeauthoryear{Semantic3d}{Semantic3d}{2016}]%
        {semantic3d}
\bibfield{author}{\bibinfo{person}{Semantic3d}.}
  \bibinfo{year}{2016}\natexlab{}.
\newblock \bibinfo{title}{Large-Scale Point Cloud Classification Benchmark}.
\newblock   (\bibinfo{year}{2016}).
\newblock
\showURL{%
\url{http://www.semantic3d.net/}}


\end{thebibliography}

\end{document}